\documentclass[english,prd,onecolumn,superscriptaddress,nofootinbib,preprintnumbers]{revtex4}
\usepackage[latin1]{inputenc}
\usepackage{graphicx}
\usepackage{amsmath,amsthm,amssymb}

\makeatletter

\usepackage{amsfonts}
\usepackage{dcolumn}
\usepackage{hyperref}

\def\be{\begin{equation}}
\def\ee{\end{equation}}
\def\ba{\begin{eqnarray}}
\def\ea{\end{eqnarray}}
\def\bs{\begin{subequations}}
\def\es{\end{subequations}}

\newcommand{\rd}{{\rm d}}

\usepackage{color}

\makeatother

\usepackage{babel}
\makeatother
\begin{document}

\title{Generalized Brans-Dicke theories}

\author{Antonio De Felice}
\affiliation{Department of Physics, Faculty of Science, Tokyo University of Science,
1-3, Kagurazaka, Shinjuku-ku, Tokyo 162-8601, Japan}

\author{Shinji Tsujikawa}
\affiliation{Department of Physics, Faculty of Science, Tokyo University of Science,
1-3, Kagurazaka, Shinjuku-ku, Tokyo 162-8601, Japan}

\begin{abstract}

In Brans-Dicke theory a non-linear self interaction of a scalar field 
$\phi$ allows a possibility of realizing the late-time cosmic acceleration, 
while recovering the General Relativistic behavior at early 
cosmological epochs. We extend this to more general modified 
gravitational theories in which a de Sitter solution for dark energy 
exists without using a field potential. We derive a condition for the stability of the 
de Sitter point and study the background cosmological 
dynamics of such theories. We also restrict the
allowed region of model parameters from the demand
for the avoidance of ghosts and instabilities. 
A peculiar evolution of the field propagation speed allows 
us to distinguish those theories from the 
$\Lambda$CDM model.

\end{abstract}

\date{\today}

\maketitle

\section{Introduction}

Scalar-tensor gravitational theories have been widely studied as 
an alternative to General Relativity (GR). In these theories 
a scalar-field degree of freedom $\phi$ is coupled to 
the Ricci scalar $R$ through a coupling of the form 
$F(\phi)R$ \cite{FujiiMaeda}.
For example, such a coupling arises in low energy effective string 
theory as a result of the dilaton coupling with gravitons \cite{GasVene}. 
The discovery of dark energy in 1998 \cite{SNIa} has also stimulated
the study for the modifications of gravity on large distances
(see Refs.~\cite{darkreview} for recent reviews).

The well-known example of scalar-tensor theories is 
Brans-Dicke (BD) theory \cite{BDtheory} 
in which $F(\phi)$ is proportional to $\phi$ with a non-canonical 
kinetic term $(-\omega_{\rm BD}/\phi)(\nabla \phi)^2$,
where $\omega_{\rm BD}$ is the so-called BD parameter.
In original BD theory without a field potential, the BD parameter 
is constrained to be $\omega_{\rm BD}>40000$ from
local gravity tests in the solar system \cite{Will}.
This comes from the fact that the coupling between the massless
field $\phi$ and non-relativistic matter needs to be suppressed
to avoid the propagation of the fifth force.
Under this bound, the deviation from GR (which corresponds 
to the limit $\omega_{\rm BD} \to \infty$) is too small to be detected
in current cosmological observations.

There are two ways to recover the General Relativistic behavior 
in high density regimes relevant to solar system experiments.
One is the so-called chameleon mechanism \cite{chame} 
in which the presence of the field potential $V(\phi)$ allows 
a possibility of having a density-dependent effective mass of the field.
Provided that the effective mass is sufficiently large in the 
regions of high density, a spherically symmetric body can have 
a thin-shell around its surface so that the effective coupling 
between $\phi$ and matter is suppressed outside the body.
This mechanism works not only for BD theory \cite{chame,TUMTY} 
but also for metric $f(R)$ gravity \cite{fRchame}, 
because the latter can be regarded as a special case of BD theory with 
$\omega_{\rm BD}=0$ \cite{ohanlon,Chiba}.
For viable dark energy models based on BD theory and metric
$f(R)$ gravity the early cosmological evolution mimics that of GR,
but the deviation from GR becomes important at late times
(i.e. low density regimes) \cite{TUMTY,fRviable,fRviable2}.

Another way for the recovery of GR in the regions of high density 
is to introduce non-linear self interactions of a scalar field, e.g., 
$\xi (\phi) \square \phi (\partial^{\mu} \phi \partial_{\mu} \phi)$, 
where $\xi$ is a function in terms of $\phi$. 
There have been attempts to restrict the form of the 
Lagrangian by imposing the ``Galilean'' symmetry 
$\partial_{\mu} \phi \to \partial_{\mu} \phi
+b_{\mu}$ \cite{Nicolis,Deffayet,Chow,Rham,Gannouji}.
The self interaction of the form 
$\square \phi (\partial^{\mu} \phi \partial_{\mu} \phi)$
respects the Galilean symmetry in the Minkowski background. 
In the Dvali-Gabadadze-Porrati (DGP) 
braneworld model \cite{DGP} the field self interaction 
$\square \phi (\partial^{\mu} \phi \partial_{\mu} \phi)$
has been employed to recover the General Relativistic 
behavior for the length scale smaller than the so-called Vainshtein 
radius \cite{DGPse1,DGPse2} (see also Refs.~\cite{Babichev}).
While this ``self screening'' mechanism (called the Vainshtein 
mechanism \cite{Vain}) can be at work for consistency with 
solar system experiments, the DGP model is unfortunately 
plagued by a ghost problem \cite{DGPse2,DGPghost} as well as 
incompatibility with a number of observational 
constraints \cite{DGPobser}.

In BD theory with the field self interaction term 
$\xi (\phi) \square \phi (\partial^{\mu} \phi \partial_{\mu} \phi)$,
there exist de Sitter (dS) solutions for $\xi(\phi) \propto \phi^{-2}$
and $\omega_{\rm BD}<-4/3$ \cite{Silva}\footnote{For constant $\xi (\phi)$
there exist no consistent dS solutions, but the presence of other field 
self interaction terms that respect Galilean symmetry 
in the Minkowski background allows a possibility for giving 
rise to dS solutions \cite{Gannouji}.}.
Moreover the General Relativistic behavior is recovered 
in early cosmological epochs during which the field is 
nearly frozen. The solutions finally approach the dS
attractor at which $\dot{\phi}/\phi$ is constant.
As long as $\dot{\phi}/\phi$ is positive, one can avoid the 
appearance of ghosts and instabilities associated with 
the field propagation speed. 
Moreover this model gives rise to several interesting 
observational signatures such as the modified growth of 
matter perturbations and anti-correlations in the cross-correlation 
of large scale structure and the integrated Sachs-Wolfe effect
in cosmic microwave background 
anisotropies \cite{Silva,Kobayashi1}.

In this paper we consider the general action (\ref{action}) below
without the field potential and derive the functional forms of 
$F(\phi)$, $B(\phi)$, and $\xi (\phi)$ from the requirement of
obtaining dS solutions with $\dot{\phi}/\phi=$\,constant.
The three functions are restricted to be of the power-law 
forms in terms of $\phi$, which include BD theory with 
$\xi (\phi) \propto \phi^{-2}$ as a special case.
We discuss the cosmological viability of such theories by 
analyzing the stability of fixed points for the background
field equations. We also study conditions for the avoidance 
of ghosts and instabilities to find the region of viable model 
parameters.

\section{Generalized Brans-Dicke theories}
\label{gbdsec}

We start with the action  
\begin{equation}
\label{action}
S=\int {\rm d}^4 x \sqrt{-g} \left[
\frac{1}{2}F(\phi)R+B(\phi)X+
\xi (\phi) \square \phi (\partial^{\mu} \phi 
\partial_{\mu} \phi) \right]
+\int {\rm d}^4 x {\cal L}_M (g_{\mu \nu}, \Psi_M)\,,
\end{equation}
where $g$ is a determinant of the metric $g_{\mu \nu}$,
$\phi$ is a scalar field with a kinetic term 
$X =-g^{\mu \nu}\partial_{\mu}\phi \partial_{\nu} \phi/2$, 
and $F(\phi)$, $B(\phi)$, $\xi (\phi)$ are functions of $\phi$.
${\cal L}_M$ is a matter Lagrangian that depends on 
the metric $g_{\mu \nu}$ and matter fields $\Psi_M$.
We would like to construct models in which 
the late-time cosmic acceleration can be realized without 
the field potential. We shall restrict the functional forms 
of $F(\phi)$, $B(\phi)$, and $\xi (\phi)$ from the 
requirement of having dS solutions.
In the presence of the nonlinear field self-interaction term
it is possible to recover the General Relativistic behavior for 
the cosmological evolution at early epochs.

We study the cosmological dynamics for a spatially flat 
Friedmann-Lema\^{i}tre-Robertson-Walker (FLRW) space-time
with the line element
\begin{equation}
{\rm d}s^2=-{\rm d}t^2+a(t)^2 \delta_{ij}
{\rm d}x^i {\rm d}x^j\,,
\label{metric}
\end{equation}
where $a(t)$ is the scale factor with cosmic time $t$.
For the matter Lagrangian ${\cal L}_M$ we consider
perfect fluids with non-relativistic matter (energy density $\rho_m$) 
and radiation (energy density $\rho_r$ and pressure $P_r=\rho_r/3$).
Varying the action (\ref{action}) with respect to $g_{\mu \nu}$
and $\phi$ for the metric (\ref{metric}), 
we obtain the following equations
\begin{eqnarray}
& & 3FH^2=B \dot{\phi}^2/2-3H F_{,\phi} \dot{\phi}+(6H \xi
-\xi_{,\phi} \dot{\phi} )\dot{\phi}^3+\rho_m+\rho_r \,,
\label{back1} \\
& & -2F \dot{H}=(F_{,\phi}-2\xi \dot{\phi}^2) \ddot{\phi}
+[ B \dot{\phi}+F_{,\phi \phi} \dot{\phi}
-HF_{,\phi}+\dot{\phi}^2 (6H\xi -2 \xi_{,\phi} 
\dot{\phi}) ] \dot{\phi} 
+\rho_m+4\rho_r/3\,,
\label{back2} \\
& & ( B -4\dot{\phi}^2 \xi_{,\phi} +12 H\xi \dot{\phi} ) 
\ddot{\phi} +[ 3HB+B_{,\phi} \dot{\phi}/2-\xi_{,\phi \phi}
\dot{\phi}^3+6 \xi (3H^2+\dot{H}) \dot{\phi}] \dot{\phi}
-3F_{,\phi} (2H^2+\dot{H})=0\,,
\label{back3}
\end{eqnarray}
where a dot represents a derivative with respect to $t$,
and $H \equiv \dot{a}/a$ is the Hubble parameter.
Note that we have used the notations
$F_{,\phi} \equiv \partial F/\partial \phi$ and 
$F_{,\phi \phi} \equiv \partial^2F/\partial \phi^2$
(the same for other variables).

{}From Eqs.~(\ref{back1}) and (\ref{back2}) it follows that 
\begin{eqnarray}
& & 1=\frac{B \phi^2}{6F}x^2-\frac{\phi F_{,\phi}}{F}x
+\frac{2 \xi \phi^3}{F} H^2 x^3 \left( 1-
\frac{\phi \xi_{,\phi}}{6 \xi} x \right)
+\frac{\rho_m}{3FH^2}+\frac{\rho_r}{3FH^2}\,,
\label{alg1}
\\
& & -2\frac{\dot{H}}{H^2}=\left( \frac{\phi F_{,\phi}}{F}
-\frac{2\xi \phi^3}{F}H^2 x^2 \right)
\left( \frac{\dot{x}}{H}+x \frac{\dot{H}}{H^2}
+x^2 \right)+\left[ \frac{B \phi^2}{F}x
+\frac{F_{,\phi \phi} \phi^2}{F}x-\frac{\phi F_{,\phi}}
{F}+\frac{6 \xi \phi^3}{F}H^2 x^2 \left(1-
\frac{\phi \xi_{,\phi}}{3\xi}x \right) \right]x \nonumber \\
& &~~~~~~~~~~~~~+
\frac{\rho_m}{F H^2}+\frac{4\rho_r}{3FH^2}\,,
\label{alg2}
\end{eqnarray}
where 
\begin{equation}
x \equiv \frac{\dot{\phi}}{H\phi}\,.
\label{xdef}
\end{equation}
Now we look for dS solutions at which $H$ and $x$
are constants. One can generalize the analysis to 
the case in which $x$ is a function of $\phi$, see 
the Appendix A.
Here we focus on theories giving constant $x$,
to recover the dS solution found in Ref.~\cite{Silva} 
as a specific case.
If $F(\phi)$ and $\xi (\phi)$ are power-law 
functions in terms of $\phi$, the quantities such as 
$\phi F_{,\phi}/F$, $F_{,\phi \phi}\phi^2/F$, and 
$\phi \xi_{,\phi}/\xi$ remain constants.
Provided that $B/F \propto \phi^{-2}$ and $\xi/F \propto \phi^{-3}$, 
one can solve Eqs.~(\ref{alg1}) and (\ref{alg2})
for $x$ and $H$ at the dS point.
From the above demands we adopt the following functions
\begin{equation}
F(\phi)=M_{\rm pl}^2 (\phi/M_{\rm pl})^{3-n}\,,\qquad
B(\phi)=\omega ( \phi/M_{\rm pl})^{1-n}\,,\qquad
\xi (\phi)=(\lambda/\mu^3) (\phi/M_{\rm pl})^{-n}\,,
\label{funchoice}
\end{equation}
where $M_{\rm pl} \simeq 10^{18}$\,GeV is 
the reduced Planck mass, $\mu~(>0)$ 
is a constant having a dimension of mass, and $\omega$
and $\lambda$ are dimensionless constants.
For the consistency of theories, the coupling $\lambda$
needs to be positive. In the Appendix B we shall show why the 
theories with $\lambda<0$ are not allowed.
The Brans-Dicke (BD) theory \cite{BDtheory} corresponds to 
$n=2$ with the BD parameter $\omega$.
Since $F(\phi)$ is constant for $n=3$, the theory with 
$n=3$ corresponds to k-essence \cite{kespapers} 
minimally coupled gravity.
Note that the choice of functional forms different from 
(\ref{funchoice}) may allow the possibility to realize 
the cosmic acceleration today, but we shall focus 
on the simplest case in which dS solutions
are the late-time attractor.

{}From Eqs.~(\ref{alg1}) and (\ref{alg2}) we obtain 
the following algebraic equations at the dS point:
\begin{eqnarray}
\omega &=& -\frac{n(n-3)^2x_{\rm dS}^3
+(n-3)(n-12)x_{\rm dS}^2-6(n-5)x_{\rm dS}
+18}{x_{\rm dS}^2 (x_{\rm dS}+3)}\, ,\label{xdS}\\
\lambda &=&\frac{\mu^3}{M_{\rm pl}H_{\rm dS}^2} 
\frac{[(n-3) x_{\rm dS}-2] 
[(n-3) x_{\rm dS}-3] }{2x_{\rm dS}^3 (x_{\rm dS}+3)}\,,
\label{HdS}
\end{eqnarray}
where $x_{\rm dS}$ and $H_{\rm dS}$ are the values of $x$ and $H$ 
at the dS point, respectively.
We fix the mass scale $\mu$ to be 
\begin{equation}
\mu = (M_{\rm pl} H_{\rm dS}^2)^{1/3} \simeq 
10^{-40}M_{\rm pl}\,,
\label{muscale}
\end{equation}
where we have used $H_{\rm dS} \simeq 10^{-60}M_{\rm pl}$.
In this case Eq.~(\ref{HdS}) is simplified as
\begin{equation}
\lambda = \frac{[(n-3) x_{\rm dS}-2] 
[(n-3) x_{\rm dS}-3] }{2x_{\rm dS}^3 (x_{\rm dS}+3)}\,,
\label{lamre}
\end{equation}
which gives $x_{\rm dS}={\cal O}(1)$
for $\lambda$ and $n$ of the order of unity.
For given $\omega$ and $n$, the quantity $x_{\rm dS}$
is determined by solving Eq.~(\ref{xdS}).
Then the dimensionless constant $\lambda$ is known from 
Eq.~(\ref{lamre}). The demand for realizing the cosmic 
acceleration today fixes the value of $\lambda$
together with the mass scale $\mu$ given in Eq.~(\ref{muscale}).

For BD theory ($n=2$), Eq.~(\ref{xdS}) can be analytically 
solved as \cite{Silva}
\begin{equation}
x_{\rm dS} =\frac{-2 \pm \sqrt{-8-6\omega}}
{2+\omega}\,,\quad -3\,.
\label{dSsol}
\end{equation}
As we will see later, the condition for the avoidance of ghosts
demands that $x>0$ during the cosmological evolution from 
the radiation era to the dS epoch.
When $n=2$, we obtain $x_{\rm dS}>0$ only for 
the solution $x_{\rm dS} =(-2-\sqrt{-8-6\omega})/(2+\omega)$
with $\omega<-2$.

In order to recover the General Relativistic behavior in the early 
cosmological epoch we require that the field initial value $\phi_i$
is close to $M_{\rm pl}$.
As we will see later, the quantity $x$ is much smaller than 1
in the early cosmological epoch. Hence the field is nearly 
frozen during the radiation and matter eras.
The field starts to evolve at the late cosmological 
epoch in which $x$ grows to the order of unity. 

For the functions (\ref{funchoice}), let us derive autonomous 
equations of the dynamical system. In addition to the quantity 
$x$ defined in Eq.~(\ref{xdef}) we also introduce 
the following variables
\begin{equation}
\label{ydef}
y \equiv \lambda x^2\frac{H^2}{H_{\rm dS}^2}\,,
\qquad
\Omega_r \equiv \frac{\rho_r}{3FH^2}\,.
\end{equation}
Since we are considering the theories with $\lambda>0$, 
the variable $y$ is positive definite.
Equation (\ref{alg1}) gives the constraint equation
\begin{equation}
\Omega_m \equiv \frac{\rho_m}{3FH^2}=
1-\Omega_r-\Omega_{\rm DE}\,,
\label{Omegamdef}
\end{equation}
where 
\begin{equation}
\Omega_{\rm DE} \equiv \frac{\omega}{6}x^2
-(3-n)x+2xy \left( 1+\frac{n}{6}x \right)\,.
\end{equation}
The variables $x$, $y$, and $\Omega_r$ 
obey the following differential equations
\begin{eqnarray}
& & x'=-\frac{d_2}{d_1}x-\frac{9x}{2\omega d_1} (3-n-2y)
\left( \Omega_m+\frac43 \Omega_r \right)+\frac{6(3-n)x}{\omega d_1}
-x^2 -x \frac{H'}{H}\,,
\label{auto1}\\
& & y'=-2y \left[ \frac{d_2}{d_1}+\frac{9}{2 \omega d_1}
(3-n-2y) \left( \Omega_m+\frac43 \Omega_r \right)
-\frac{6(3-n)}{\omega d_1}+x \right]\,,
\label{auto2}\\
& & \Omega_r'=-2\Omega_r \left( 2+\frac{3-n}{2}x+ \frac{H'}{H} \right)\,,
\label{auto3}
\end{eqnarray}
where a prime represents a derivative with respect to $N=\ln a$, and 
\begin{eqnarray}
\hspace*{-1.0em}
& & d_1 \equiv \frac{12}{\omega}y+x \left[ 1+\frac{4n}{\omega}y
+\frac{3}{2\omega} (3-n-2y)^2 \right]\,,\\
\hspace*{-1.0em}
& & d_2 \equiv \frac{18}{\omega}y+x \biggl\{ 3+\frac{1-n}{2}x
-\frac{n(n+1)}{\omega}xy+3(3-n-2y) \biggl[ \frac{x}{2}
+\frac{(3-n)(2-n)}{2\omega}x-\frac{3-n}{2\omega}
+\frac{3}{\omega}y \left( 1+\frac{n}{3}x \right) \biggr] \biggr\}.
\end{eqnarray}
The Hubble parameter satisfies the equation
\begin{eqnarray}
\frac{H'}{H} &=& \frac12 (3-n-2y) \left[ \frac{d_2}{d_1}x
+\frac{9x}{2 \omega d_1} (3-n-2y) \left( \Omega_m
+\frac43 \Omega_r \right)-\frac{6(3-n)}{\omega d_1}x \right]
\nonumber \\
& & -\frac12 x \left[ \omega x+(3-n)(2-n)x-(3-n)+
y (6+2nx) \right]-\frac32 \Omega_m-2\Omega_r\,.
\label{doth}
\end{eqnarray}
We also define the effective equation of state 
\begin{equation}
w_{\rm eff} \equiv -1-\frac{2H'}{3H}\,.
\label{weff}
\end{equation}
The relation (\ref{doth}) should be substituted into 
Eqs.~(\ref{auto1}), (\ref{auto3}), and (\ref{weff})
to solve the cosmological dynamics numerically.
At the dS point ($H=$\,constant) we have that $w_{\rm eff}=-1$.

In addition to the dS point derived above, there is a fixed 
point for the system (\ref{auto1})-(\ref{auto3}) that corresponds
to the matter-dominated epoch:
\begin{equation}
P_m:~(x, y, \Omega_r)=\left(0, \frac{3-n}{6}, 0 \right)\,.
\label{matter}
\end{equation}
Note that this can be also derived by equating the term 
$6\xi (3H^2+\dot{H})\dot{\phi}^2$
in Eq.~(\ref{back3}) with the term $3F_{,\phi}(2H^2+\dot{H})$.
For $\lambda$ and $n$ of the order of unity we have that 
$x_{\rm dS}={\cal O}(1)$.
{}From the definition of $y$ given in 
Eq.~(\ref{ydef}) we have $x^2 \approx y (H_{\rm dS}^2/H^2) \ll y$
during the matter and the radiation eras.
Since $y$ is positive definite, we require the condition  
\begin{equation}
n \le 3\,.
\label{ncon}
\end{equation}
Considering linear perturbations $\delta x$, $\delta y$, $\delta \Omega_r$
about the fixed point (\ref{matter}), the eigenvalues of the Jacobian 
matrix of perturbations are given by  $3/2$, $-3/2$, $-1/2$.
Hence the matter point is a saddle followed by the dS solution.

There are no fixed points that correspond to the radiation-dominated epoch.
However one can analytically estimate the evolution of the variable 
$y$ under the condition that $x$ is negligibly small relative to $y$.
Using the approximation that $d_1 \simeq 12 y/\omega$,  
$d_2 \simeq 18 y/\omega$, and $\Omega_r \simeq 1-\Omega_m$, 
we obtain the following equation during radiation 
and deep matter eras:
\begin{equation}
y' \simeq -y+\frac14 (3-n-2y)\Omega_m\,.
\label{yeq}
\end{equation}
Provided that $2y \ll 3-n$, the solution to
Eq.~(\ref{yeq}) during the radiation domination is given by 
\begin{equation}
y \simeq \frac18 (3-n) \Omega_m+ce^{-N}\,,
\label{yestima}
\end{equation}
where $c$ is a constant. To derive the solution (\ref{yestima})
we have used the relation $\Omega_m'=\Omega_m \propto a$.
Since the term $ce^{-N}$ decays with time, one has
$y \simeq (3-n)\Omega_m/8 \propto a$ for $2y \ll 3-n$ and $n \neq 3$.
When the term $2y$ grows to the same order as
$3-n$ during the matter dominance ($\Omega_m \simeq 1$)
one can obtain the matter point (\ref{matter}) 
by setting $y'=0$ in Eq.~(\ref{yeq}).
Note that if $n=3$ the variable $y$ decreases as $y \propto e^{-N}$ 
during the radiation domination.

The above discussion shows that the radiation era, characterized
by $(x, y, \Omega_r) \simeq (0, (3-n)\Omega_m/8, 1)$ and 
$w_{\rm eff}=1/3$, is followed by the matter era
with $(x, y, \Omega_r) \simeq (0, (3-n)/6, 0)$ and $w_{\rm eff}=0$.
The solutions finally approach the dS point characterized
by the conditions (\ref{xdS}) and (\ref{HdS}), provided that 
it is stable. In the next section we derive the stability 
condition for the dS point by considering homogeneous
perturbations about it. We also restrict the viable region of
model parameters from the requirement to avoid ghosts 
and instabilities.

\section{Stability of the de Sitter solution, ghosts, and instabilities}
\label{persec}

In order to discuss the stability of the dS solution as well as
conditions for the avoidance of ghosts and instabilities, we consider 
scalar metric perturbations $\alpha$, $\beta$, $\psi$, and $\gamma$
about the flat FLRW metric \cite{Bardeen}
\begin{equation}
{\rm d}s^2=-(1+2\alpha){\rm d}t^2-2a(t)\partial_i \beta\,
{\rm d}t {\rm d}x^i+a^2(t) \left( \delta_{ij}-2\psi \delta_{ij}
+2\partial_i \partial_j \gamma \right) {\rm d}x^i {\rm d}x^j\,.
\label{metric2}
\end{equation}
We introduce the gauge-invariant curvature perturbation \cite{Lukash}
\begin{equation}
{\cal R} \equiv \psi +\frac{H}{\dot{\phi}} \delta \phi\,,
\end{equation}
which can be used to discuss the stability of cosmological solutions.

\subsection{Stability of the dS solution}

We expand the action (\ref{action}) at second order in the perturbation around the dS
background. 
There is only one propagating scalar degree of freedom, which
corresponds to the curvature perturbation ${\cal R}$.
In terms of the gauge invariant quantity ${\cal R}$,
the second-order perturbed action about the dS point is given by 
\begin{equation}
\label{eq:actdeSp}
\delta S^{(2)}=\int \rd t\,\rd^3 x a^3\,Q_s \left[\frac12 (\partial_t {\cal R})^2
-\frac12 \frac{c_s^2}{a^2}(\nabla {\cal R})^2 \right]\,,
\end{equation}
where 
\begin{eqnarray}
Q_s &\equiv& \frac{2F\,\Gamma}
{(F_{,\phi} \dot{\phi}+2H_{\rm dS}F-2\xi \dot{\phi}^3)^2}\,,
\label{Qsdef} \\
c_s^2 &\equiv& \frac{1}{\Gamma (F_{,\phi}-2\xi \dot\phi^2)}
 \biggl[ 8 \xi^3 \dot\phi^8+4(F_{,\phi} \xi -2F \xi_{,\phi})
 \xi \dot{\phi}^6+64 H_{\rm dS} F \xi^2 \dot{\phi}^5-2(5 F_{,\phi}^2 \xi +
4 F F_{,\phi \phi} \xi -2 F F_{,\phi}\xi_{,\phi} ) \dot{\phi}^4 \nonumber \\
& &-72 H_{\rm dS} F F_{,\phi}\xi \dot\phi^3 -3(24 \xi H_{\rm dS}^2 F^2
-F_{,\phi}^3) \dot{\phi}^2+12H_{\rm dS} F F_{,\phi}^2 \dot{\phi}
+12H_{\rm dS}^2 F^2 F_{,\phi} \biggr]\,,
\label{csdef}
\end{eqnarray}
and
\begin{equation}
\Gamma \equiv 12 \xi^2 \dot{\phi}^6-4 (3F_{,\phi} \xi+
F \xi_{,\phi} )\dot{\phi}^4+3F_{,\phi}^2 \dot{\phi}^2
+12H_{\rm dS}FF_{,\phi} \dot{\phi}
+12H_{\rm dS}^2 F^2\,.
\end{equation}
The scale factor evolves as $a\propto \exp(H_{\rm dS}\,t)$ 
at the dS point.
If $\xi=0$, then the propagation speed reduces to unity. 

For the power-law functions of $F(\phi)$, $B(\phi)$, and $\xi (\phi)$
given in Eq.~(\ref{funchoice}), Eqs.~(\ref{Qsdef}) and 
(\ref{csdef}) reduce to 
\begin{eqnarray}
Q_s &=& \frac{2 \kappa^{1-n} \phi^{3-n}\,[(n-3)(3n^2-10n+12)x_{\rm dS}^3
+12(n-5)x_{\rm dS}^2+18(n-8)x_{\rm dS}-108]}
{(n-2)^2\, [(n-3)x_{\rm dS}-2]\, x_{\rm dS}^2}\,,
\label{eq:QdS}\\
c_s^2 &=&-\frac{(n-2) [(n-3)(n-4)x_{\rm dS}^2-8(n-3)x_{\rm dS}+6]
x_{\rm dS}}{(n-3)(3n^2-10n+12)x_{\rm dS}^3
+12(n-5)x_{\rm dS}^2+18(n-8)x_{\rm dS}-108}\,.
\label{eq:csdS}
\end{eqnarray}
{}From the action (\ref{eq:actdeSp}) we obtain the equation for 
the curvature perturbation in Fourier space:
\begin{equation}
\frac{1}{a^3 Q_s} \frac{{\rm d}}{{\rm d}t}
(a^3 Q_s \dot{\cal R})+c_s^2 \frac{k^2}{a^2}
{\cal R}=0\,,
\label{Qseq}
\end{equation}
where $k$ is a comoving wavenumber.
For homogeneous perturbations ($k=0$) which have only the time-dependence,
the solution to Eq.~(\ref{Qseq}) is given by
\begin{equation}
{\cal R}(t)=c_1+c_2 \int \frac{1}{a^3 Q_s}\,{\rm d}t\,,
\end{equation}
where $c_1$ and $c_2$ are integration constants.
Since $x_{\rm dS}$ is constant at the dS point one has 
$\phi \propto \exp(x_{\rm dS}H_{\rm dS}t)$ and hence
$Q_s \propto \exp [(3-n)x_{\rm dS}H_{\rm dS}t]$ from 
Eq.~(\ref{eq:QdS}). Then the homogeneous perturbation
about the dS point evolves as 
\begin{equation}
{\cal R}(t)=c_1+\tilde{c}_2 
\exp\{[(n-3)x_{\rm dS} -3] H_{\rm dS}\,t\}\,,
\end{equation}
where $\tilde{c}_2$ is a constant.
In order to avoid the growth of ${\cal R}$, one requires that 
\begin{equation}
(n-3) x_{\rm dS}<3\,,
\label{dScon}
\end{equation}
which corresponds to the stability condition for the dS point.
Note that the same condition follows by considering the stability of 
perturbations $\delta x$, $\delta y$, and $\delta \Omega_r$
for Eqs.~(\ref{auto1})-(\ref{auto3}).
Since $n \le 3$, Eq.~(\ref{dScon}) is automatically satisfied 
for $x_{\rm dS}>0$. In fact, the variable $x$ needs to be positive 
during the cosmological evolution  starting from the radiation era to avoid
the appearance of ghosts. In the following we will show 
how this condition arises.

\subsection{Conditions for the avoidance of ghosts and instabilities}

Let us derive conditions for the avoidance of ghosts and instabilities by taking 
into account non-relativistic matter (equation of state $w_m \simeq +0$)
and radiation (equation of state $w_r \simeq 1/3$).
The velocity potentials $v_m$ and $v_r$ are related with the energy-momentum 
tensors ${T^{0}_{i}}^{(m)}$ and ${T^{0}_{i}}^{(r)}$ of 
non-relativistic matter and radiation, respectively, as 
${T^{0}_{i}}^{(m)}=-\rho_m v_{m,i}$ and 
${T^{0}_{i}}^{(r)}=-(\rho_r+P_r) v_{r,i}$. 
We introduce the following gauge-invariant combinations
\begin{equation}
V_m \equiv v_m-\frac{\delta\phi}{\dot\phi}\,,\qquad
V_r \equiv v_r-\frac{\delta\phi}{\dot\phi}\,.
\end{equation}
Perturbing the action (\ref{action}) at second order, it follows that  
\begin{equation}
 \delta S^{(2)}=\frac12\int\rd t\, \rd^3x\, a^3\left[\dot{\vec V}^t 
 A\dot{\vec V}-\nabla\vec V^t P\nabla\vec V-\vec V^t B\dot{\vec V}
 -\vec V^t M\vec V\right]\,,
 \label{deltaS}
\end{equation}
where $\vec V=({\cal R}, V_m, V_r)$, and $A$, $P$, $B$, $M$ are time-dependent 
$3 \times 3$ matrices. Here we do not write the explicit forms of the matrices 
because of their complexities. The readers may refer to the papers \cite{DeFelice,DMT}
for the details of such studies.

The ghost conditions can be derived by the signs of the 
eigenvalues of $A$.
Diagonalizing the 3$\times$3 matrix $A$ with components $A_{ij}$, 
we find that the ghosts are absent under the conditions
\begin{equation}
{\rm det}\,A>0\,,\qquad 
A_{22}A_{33}-A_{23}^2>0\,, \qquad
A_{33}>0\,.
\end{equation}
In the uniform field gauge given by $\delta\phi=0$, these 
conditions translate into
\begin{eqnarray}
\hspace{-2.5em}& &\det A=\frac{\rho_m(1+w_m)}{w_m} \frac{\rho_r(1+w_r)}{w_r} 
\frac{2F\dot{\phi}^2 [3F_{,\phi}^2+2FB+12\xi^2\dot{\phi}^4
-4(2F \xi_{,\phi}+3F_{,\phi}\xi )\dot{\phi}^2+24HF \xi \dot{\phi}]}
{(F_{,\phi} \dot\phi+2 F H-2 \xi \dot\phi^3)^2}>0\,,
\label{ghost1}\\
\hspace{-2.5em}& & A_{22}A_{33}-A_{23}^2=\frac{\rho_m(1+w_m)}{w_m} 
\frac{\rho_r(1+w_r)}{w_r}>0\,, 
\label{ghost2}\\
\hspace{-2.5em}& & A_{33}=\frac{\rho_r(1+w_r)}{w_r}>0
\label{ghost3}\,.
\end{eqnarray}
Since the conditions (\ref{ghost2}) and (\ref{ghost3}) automatically hold, 
the no-ghost condition is simply given by ${\rm det}~A>0$, i.e., 
\begin{equation}
Q_s \equiv \frac{2F\dot{\phi}^2 [3F_{,\phi}^2+2FB+12\xi^2\dot{\phi}^4
-4(2F \xi_{,\phi}+3F_{,\phi}\xi )\dot{\phi}^2+24HF \xi \dot{\phi}]}
{(F_{,\phi} \dot\phi+2 F H-2 \xi \dot\phi^3)^2}>0\,.
\label{Qdef}
\end{equation}
Eliminating the term $B$ by using the background equation (\ref{back1}),
the above definition of $Q_s$ is identical 
to (\ref{Qsdef}) at the dS point.
For the power-law functions given in Eq.~(\ref{funchoice}) this
condition reduces to 
\begin{equation}
Q_s=\frac{2F x [x(27-18n+3n^2+2\omega+12y^2+20ny-36y)
+24y]}{[x(n-3+2y)-2]^2}>0\,.
\end{equation}
During the radiation and matter eras one has $|x| \ll y \lesssim {\cal O}(1)$
and hence
\begin{equation}
Q_s \simeq 12Fxy>0 \quad \to 
\quad Fx>0\,.
\label{xcon}
\end{equation}
The no-ghost condition arising from vector and tensor perturbations
corresponds to $F>0$. 
We then find that Eq.~(\ref{xcon}) gives $x=\dot{\phi}/(H\phi)>0$. 
As long as $\phi>0$ initially the field $\phi$ and the function 
$F(\phi)=M_{\rm pl}^2 (\phi/M_{\rm pl})^{3-n}$ grow for $n<3$, 
which means that $F(\phi)$ remains positive.
If the variable $x$ crosses 0 from positive to negative, this corresponds 
to the violation of the no-ghost condition because $Q_s$ becomes 
negative just after the crossing.
Hence we require that $x>0$ during the cosmological evolution starting from
the radiation era.
Under the condition (\ref{ncon}) together with $F>0$ and $x>0$, 
we find that $Q_s$ in Eq.~(\ref{eq:QdS}) is positive at the dS point.

Let us next discuss the three speeds of propagation. 
Two of them are trivial, $c_m^2=w_m$, $c_r^2=w_r$, 
whereas the third one can be written as
\begin{eqnarray}
c_s^2 &=&\frac{1}{Z^2}\,\biggl\{-48 \xi^4 \dot\phi^{12}+64 F \xi^2 \xi_{,\phi} 
\dot\phi^{10}-384 HF \xi^3 \dot\phi^9 +[(72 F_{,\phi}^2+48 F F_{,\phi\phi}) \xi^2+
(-32 F F_{,\phi} \xi_{,\phi}+16 F^2 \xi_{,\phi\phi}) \xi  \nonumber \\
& &~~~~~~~-16 F^2 \xi_{,\phi}^2] \dot\phi^8
+(-64 H F^2  \xi \xi_{,\phi}+576H F F_{,\phi} \xi^2) \dot\phi^7
+[(672 F^2 H^2
+((48 w_m-80) \rho_m+(48 w_r-80) \rho_r) F) \xi^2
\nonumber \\
& &~~~~~~~-(24 FF_{,\phi}F_{,\phi\phi}  
+8 F^2 B_{,\phi}+48 F_{,\phi}^3) \xi] \dot\phi^6
-336 H F F_{,\phi}^2 \xi \dot\phi^5 
+[(-624 H^2 F^2 F_{,\phi}+((88-24 w_m) \rho_m
\nonumber \\
& &~~~~~~~+(88-24 w_r) \rho_r) F F_{,\phi} ) \xi
+9 F_{,\phi}^4] \dot\phi^4 
+[(-384 H^3 F^3+128(\rho_m +\rho_r )H F^2) \xi
+72 H F F_{,\phi}^3] \dot\phi^3 \nonumber \\
& &~~~~~~~+[216 H^2 F^2 F_{,\phi}^2 
-24(\rho_m+\rho_r) F F_{,\phi}^2] \dot\phi^2 
+[288 H^3 F^3 F_{,\phi}-96(\rho_m+ \rho_r)H  
F^2 F_{,\phi}] \dot\phi
+144 H^4 F^4 \nonumber \\
& &~~~~~~~-96(\rho_m+\rho_r)H^2 F^3 
+16(\rho_m+\rho_r)^2 F^2 \biggr\}\,,
\label{cses}
\end{eqnarray}
where
\begin{equation}
Z \equiv \dot{\phi}^2 [3F_{,\phi}^2+2FB+12\xi^2\dot{\phi}^4
-4(2F \xi_{,\phi}+3F_{,\phi}\xi )\dot{\phi}^2+24HF \xi \dot{\phi}]\,.
\end{equation}
We require that $c_s^2>0$ to avoid the instabilities for small-scale
perturbations.

Plugging $w_m=0$ and $w_r=1/3$ into Eq.~(\ref{cses}) and using 
$\Omega_r$ and $\Omega_m$ defined in Eqs.~(\ref{ydef}) and 
(\ref{Omegamdef}), it follows that 
\begin{align}
c_s^2&=\{[-48 y^4+16 n y^3+(-824 n+192 n^2+936+40 \omega) y^2+
(-1728+68 n \omega-140 \omega-768 n^2+96 n^3+2016 n) y\notag\\
&\qquad-72 n \omega+4 \omega^2-108 n^3+12 n^2 \omega+
729-972 n+108 \omega+486 n^2+9 n^4] x^2+[96 y^3+(-576+320 n) y^2\notag\\
&\qquad+(32 \omega+648-432 n+72 n^2) y] x
+48(\Omega_r+5) y^2+24(3-n)(1-\Omega_r) y\}\notag\\
&\qquad{}\times[(12 y^2+(20 n-36) y+3 n^2+2 \omega+27-18 n) x
+24 y]^{-2}\,.
\label{cses2}
\end{align}
Let us estimate the evolution of $c_s^2$ during the radiation and deep matter eras
in which $x$ is negligibly small relative to $y$.
Using the relation $\Omega_r \simeq 1-\Omega_m$, we obtain
\begin{equation}
c_s^2 \simeq \frac12 +\frac{(3-n-2y)\Omega_m}{24 y}\,.
\label{csasy}
\end{equation}
During the radiation domination the variable $y$ evolves as 
$y \simeq (3-n)\Omega_m/8$ for $n \neq 3$, see Eq.~(\ref{yestima}).
Plugging this relation into Eq.~(\ref{csasy}), we find
\begin{equation}
c_s^2 \simeq \frac{5}{6}-\frac{1}{12} \Omega_m\approx\frac56\,,\qquad
({\rm radiation~era~for}~n \neq 3)\,,
\label{csra}
\end{equation}
which is valid for $\Omega_m \ll 1$.
During the matter dominance the evolution of $y$ is given by $y=(3-n)/6$.
{}From Eq.~(\ref{csasy}) we then obtain 
\begin{equation}
c_s^2 \simeq \frac{2}{3}\,,\qquad
({\rm matter~era~for}~n \neq 3)\,,
\label{csma}
\end{equation}
which agrees with the result derived for $n=2$ \cite{Silva}.
This propagation speed is independent of the values of $n$.
As we will see in the next section, $c_s^2$ starts to deviate from 
$2/3$ around the late epoch of the matter era.
The above discussion shows that the instabilities associated
with the negative propagation speed squared can be avoided
during the radiation and matter eras.

While the results (\ref{csra}) and (\ref{csma}) have been derived 
for $n \neq 3$, the situation is different for $n=3$.
Plugging $n=3$ into Eq.~(\ref{csasy}), the propagation speed squared 
in the regime $x \ll y$ is given by  
\begin{equation}
c_s^2 \simeq \frac{1}{2}-\frac{1}{12} \Omega_m\,,\qquad
({\rm early~cosmological~epoch~for}~n=3)\,.
\label{csra2}
\end{equation}
As we see in the next section, this estimation ceases to be valid 
during the matter era because $x$ and $y$ become 
the same order for $n=3$.

\begin{figure}
\begin{centering}
\includegraphics[width=4.0in,height=3.7in]{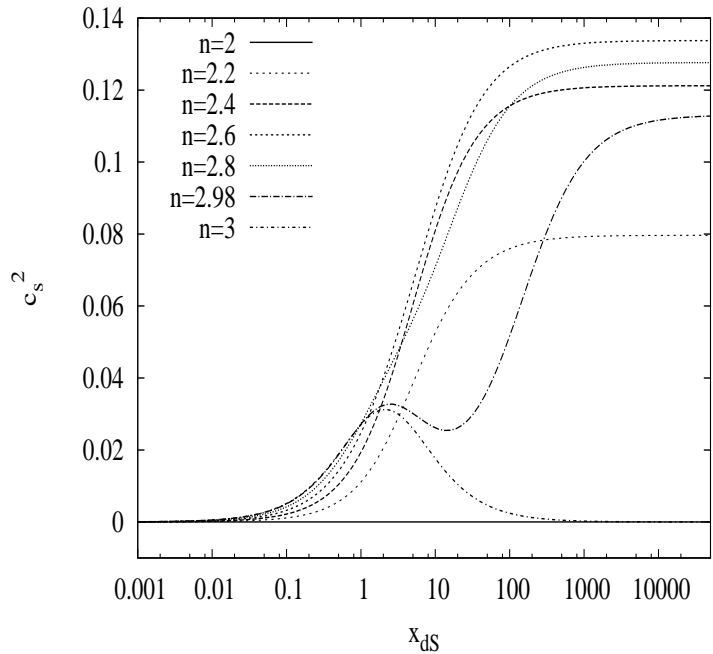} 
\par\end{centering}
\caption{The propagation speed squared at the dS point 
versus $x_{\rm dS}$ for several different values of $n$. 
One has $0 \le c_s^2<1$ for $2 \le n \le 3$.}
\centering{}\label{csfig} 
\end{figure}

At the dS fixed point, $c_s^2$ is given by 
Eq.~(\ref{eq:csdS}). Under the conditions $n \le 3$
and $x_{\rm dS}>0$, one has $c_s^2 \ge 0$ for 
\begin{equation}
n \ge 2\,.
\label{ncon2}
\end{equation}
In Fig.~\ref{csfig} we plot $c_s^2$ at the dS point 
as a function of $x_{\rm dS}$ for a number of
different values of $n$.
One has $c_s^2=0$ for $n=2$.
In the regime $x_{\rm dS} \ll 1$ the propagation speed squared
is approximately given by $c_s^2 \simeq (n-2)x_{\rm dS}/18$.
If $2<n<3$ the asymptotic value of $c_s^2$ in the limit 
$x_{\rm dS} \to \infty$ is
\begin{equation}
\lim_{x_{\rm dS} \to \infty} c_s^2
=\frac{(n-2)(4-n)}{3n^2-10n+12}\,,
\end{equation}
which has the maximum value $c_s^2=0.134$ at $n=2.618$.
In Fig.~\ref{csfig} we find that, for  $n \lesssim 2.9$, 
$c_s^2$ monotonically increases with $x_{\rm dS}$.
The local extrema of $c_s^2$ appear for $n$ close to 3.
When $n=3$ the propagation speed squared has a 
dependence $c_s^2 \simeq 1/(4x_{\rm dS})$ for $x_{\rm dS} \gg 1$
and hence $c_s^2$ approaches 0 in the limit that 
$x_{\rm dS} \to \infty$.
{}From the above discussion we have $0 \le c_s^2<0.134$ for 
$2 \le n \le 3$.

\subsection{Viable parameter space}

Let us summarize the viable region of model parameters.
{}From Eqs.~(\ref{ncon}) and (\ref{ncon2}) the parameter $n$
is restricted in the range
\begin{equation}
2 \le n \le 3\,.
\label{nfin}
\end{equation}
The condition $2 \le n$ comes from the requirement $c_s^2 \ge 0$
at the dS point, whereas another condition $n \le 3$ is needed
for the existence of the matter fixed point.
Provided $F>0$, we require that
\begin{equation}
x>0\,,
\label{xfin}
\end{equation}
to avoid the appearance of ghosts during the 
cosmological evolution from the radiation era to the dS epoch.
Under (\ref{nfin}) and (\ref{xfin}) the condition (\ref{dScon})
for the stability of the dS point is automatically satisfied.

\begin{figure}
\begin{centering}
\includegraphics[width=3.7in,height=3.6in]{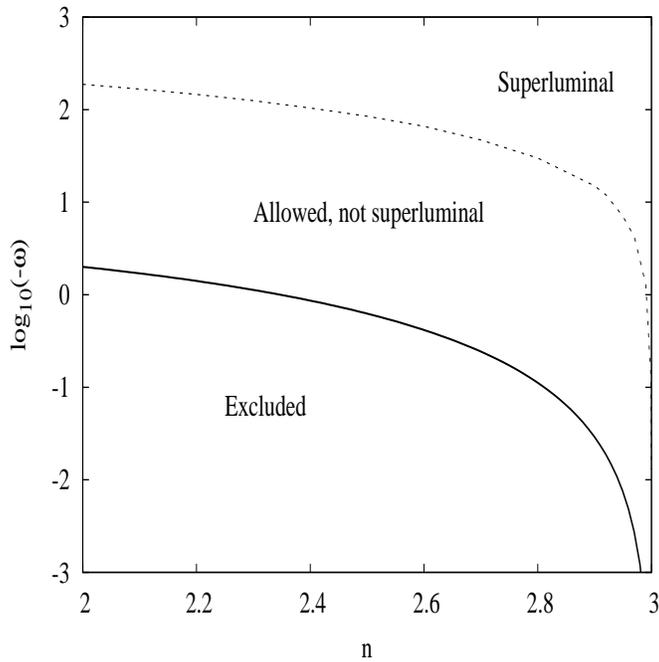} 
\par\end{centering}
\caption{The allowed and excluded regions in the 
$(n, \omega)$ plane. The vertical axis is plotted in 
terms of $\log_{10} (-\omega)$ with $\omega<0$.
The allowed region corresponds to 
$2 \le n \le 3$ and $\omega<-n(n-3)^2$.
The black line shows $\omega=-n(n-3)^2$, whereas
the dotted line correspond to the border at which 
the propagation speed squared temporally 
reaches $c_s^2=1$ during the course of 
the cosmological evolution.}
\centering{}\label{omegafig} 
\end{figure}

{}From Eq.~(\ref{xdS}) one can restrict the parameter region 
of $\omega$ under the conditions (\ref{nfin}) and (\ref{xfin}).
In the limit that $x_{\rm dS} \to \infty$ the asymptotic 
value of $\omega$ is $\omega_{\infty}=-n(n-3)^2$.
Meanwhile, in the limit that $x_{\rm dS} \to 0$, we have 
$\omega \simeq  -6/x_{\rm dS}^2 \to -\infty$.
The viable dS point exists for 
\begin{equation}
\omega<-n(n-3)^2\,,
\label{wcon}
\end{equation}
which does not have a lower bound.
When $n=2$ and $n=3$ the allowed region 
corresponds to $\omega<-2$ and $\omega<0$, 
respectively. 
In Fig.~\ref{omegafig} we illustrate the viable parameter
space in the $(n, \omega)$ plane. The black line 
represents the border characterized by $\omega=-n(n-3)^2$.

Although we have shown that $c_s^2$ exists in the regime 
$0 \le c_s^2 <1$ during the radiation, matter, and dS epochs
under the conditions (\ref{nfin}) and (\ref{xfin}), 
it can happen that the speed of propagation temporally reaches 
the superluminal regime ($c_s^2>1$) during the transition from 
the matter era to the dS epoch.
We shall discuss about this in the next section.

\section{Cosmological evolution}
\label{background}

In this section we shall numerically integrate the dynamical equations 
(\ref{auto1})-(\ref{auto3}) to confirm the analytic estimation 
given in previous sections.
When $n \neq 3$ the variable $y$ evolves as 
$y \simeq (3-n)\Omega_m/8 \propto a$ 
during the radiation era, which is followed by the matter era with 
$y \simeq (3-n)/6$\,=\,constant.
Finally the solutions approach the dS point with 
\begin{equation}
y_{\rm dS} =\frac{[(n-3)x_{\rm dS}-2][(n-3)x_{\rm dS}-3]}
{2x_{\rm dS} (x_{\rm dS}+3)}\,.
\label{ysolu}
\end{equation}
For $n$ and $\omega$ satisfying the conditions 
(\ref{nfin}) and (\ref{wcon}), $x_{\rm dS}~(>0)$ is known 
by solving Eq.~(\ref{xdS}). 

\begin{figure}
\begin{centering}
\includegraphics[height=3.3in,width=3.3in]{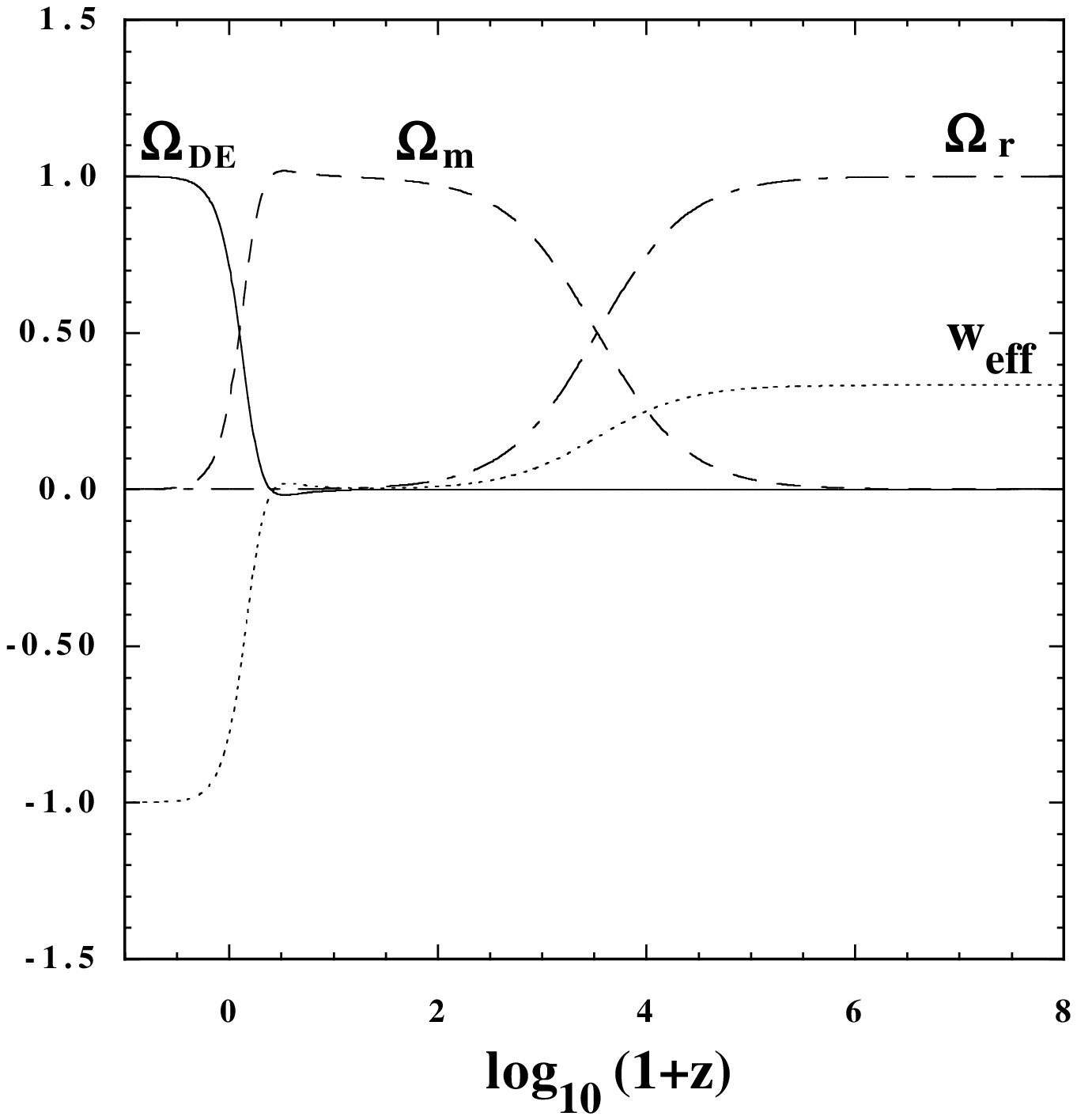}
\includegraphics[height=3.2in,width=3.3in]{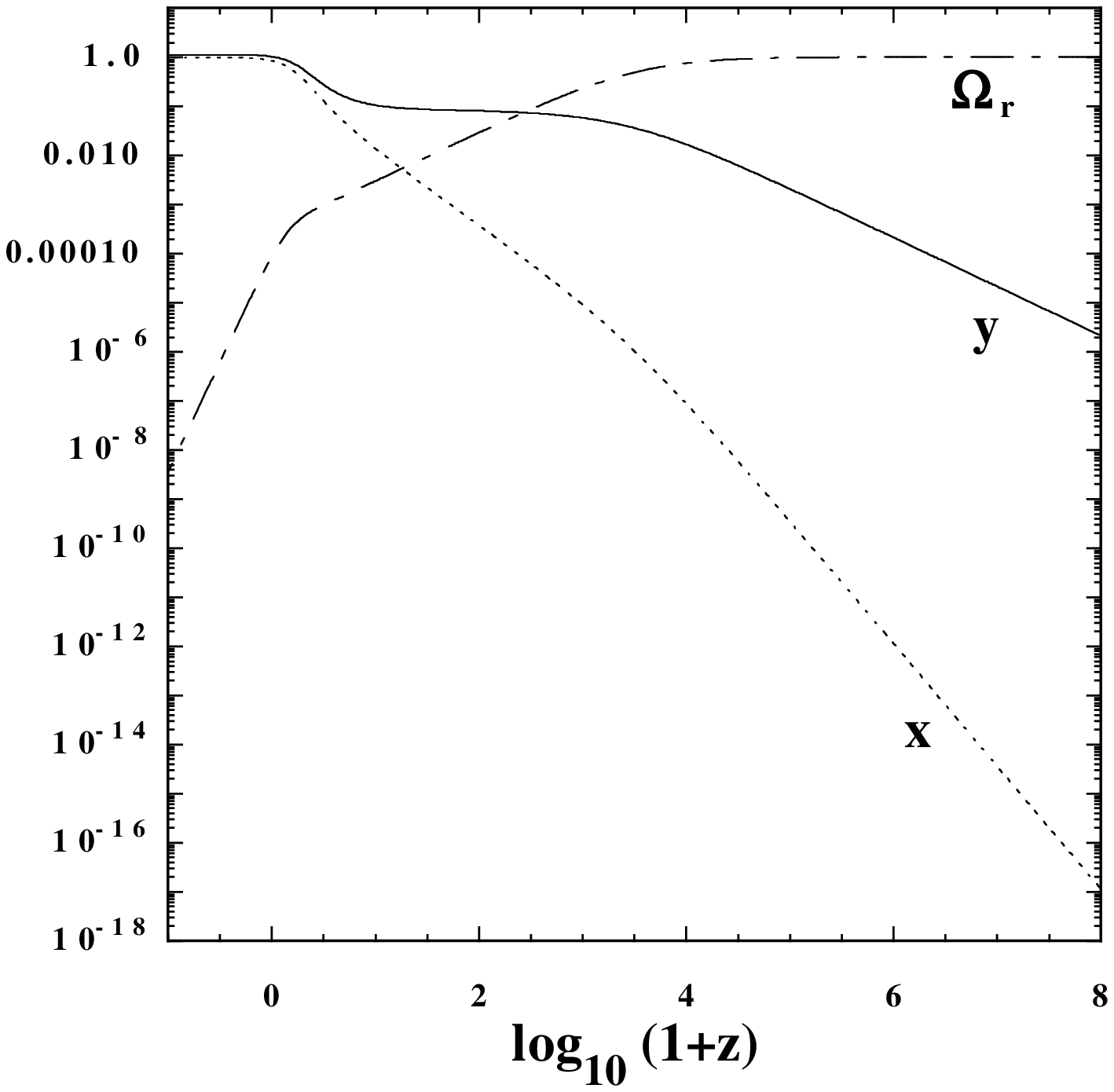}  
\par\end{centering}
\caption{Evolution of $\Omega_{\rm DE}$, $\Omega_m$, $\Omega_r$, 
and $w_{\rm eff}$ versus the redshift $z=1/a-1$ 
for $n=2.5$ and $\omega=-10$ (left panel). 
The initial conditions are chosen to be 
$x=10^{-18}$ and $y=(3-n)\Omega_m/8$ with $\Omega_m=1.28 \times 10^{-5}$.
We identify the present epoch ($z=0$) as $\Omega_{\rm DE}=0.72$, 
$\Omega_m=0.28$, and $\Omega_r=8 \times 10^{-5}$. 
The right panel shows the evolution of the variables $x$, $y$ 
and $\Omega_r$ with a logarithmic scale for the same 
model parameters and initial conditions
as those in the left panel. }
\label{backevo}
\end{figure}

In Fig.~\ref{backevo} we plot the evolution of $\Omega_{\rm DE}$, 
$\Omega_m$, and $\Omega_r$ as well as $x$ and $y$ versus 
the redshift $z$ for $n=2.5$ and $\omega=-10$. 
In this case Eqs.~(\ref{xdS}) and (\ref{ysolu}) give 
$x_{\rm dS}=0.974$ and $y_{\rm dS}=1.120$, 
which coincide with the numerical values at the dS point.
The numerical evolution of the variable $y$ shown 
in the right panel of Fig.~\ref{backevo} is consistent with 
the analytic estimation of $y$ for each cosmological epoch.
The evolution of the variable $x$ during radiation and matter eras
can be understood as follows. Using the relations 
$x \ll y \lesssim {\cal O}(1)$, $\Omega_r \simeq 1-\Omega_m$, 
and $H'/H \simeq -3\Omega_m/2-2\Omega_r$ in these epochs, 
Eq.~(\ref{auto1}) reduces to 
\begin{equation}
x' \simeq x \left[ \frac32 +\frac{(3-n)\Omega_m}{8y}
-\frac34 \Omega_m \right]\,.
\end{equation}
Since $y \simeq (3-n)\Omega_m/8$ for $n \neq 3$ during 
the radiation dominance, we obtain the solution $x \propto e^{5N/2}=a^{5/2}$.
During the matter dominance characterized by $y \simeq (3-n)/6$ and 
$\Omega_m \simeq 1$, the solution is given by 
$x \propto e^{3N/2}=a^{3/2}$.
Hence the variable $x$ evolves faster than $y$, 
as we see in the right panel of Fig.~\ref{backevo}.

The left panel of Fig.~\ref{backevo} shows that the successful sequence
of radiation ($w_{\rm eff} \simeq 1/3$, $\Omega_r \simeq 1$), 
matter ($w_{\rm eff} \simeq 0$, $\Omega_m \simeq 1$), and
dS eras ($w_{\rm eff} \simeq -1$, $\Omega_{\rm DE} \simeq 1$) are 
in fact realized. Even if we start integrating from the high-redshift regime
the solutions are not prone to numerical instabilities, unlike modified 
gravity models having a large mass in the regions of high density
(such as $f(R)$ dark energy models \cite{fRviable2,Appleby,Moraes}).
This property is associated with the fact that, instead of the field potential,  
the scalar-field self interaction is used to recover the General Relativistic 
behavior at early times. 

\begin{figure}
\begin{centering}
\includegraphics[width=3.6in,height=3.5in]{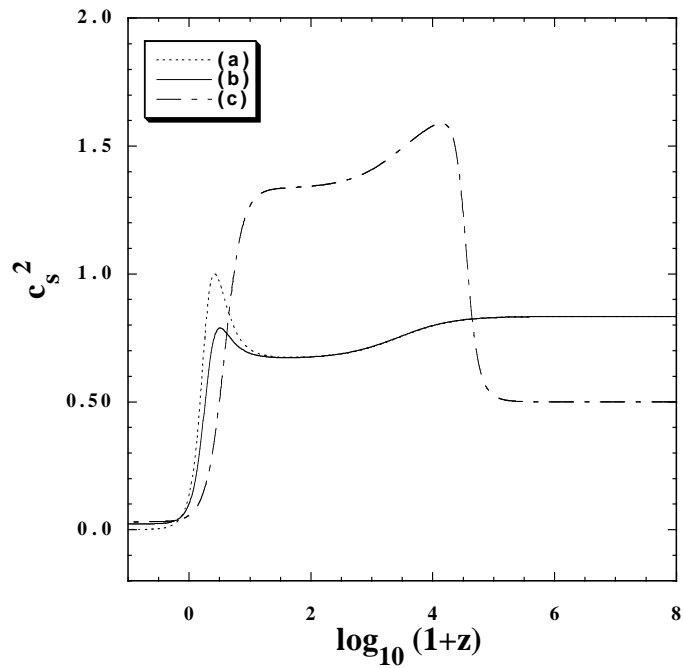} 
\par\end{centering}
\caption{Evolution of the propagation speed squared 
$c_s^2$ versus the redshift $z$ for
the cases (a) $n=2$, $\omega=-188$, 
(b) $n=2.5$, $\omega=-10$, and 
(c) $n=3$, $\omega=-1$.
The present epoch ($z=0$) is identified as
$\Omega_{\rm DE}=0.72$, $\Omega_m=0.28$, 
and $\Omega_r=8 \times 10^{-5}$. }
\centering{}\label{csevofig} 
\end{figure}

We have run numerical simulations for other model parameters 
satisfying the conditions (\ref{nfin}) and (\ref{wcon})
and confirmed that the successful cosmological evolution can 
be realized at the background level.
We have also followed the evolution of the quantity 
$F=M_{\rm pl}^2 (\phi/M_{\rm pl})^{3-n}$ by numerically 
solving the equation 
\begin{equation}
F'=(3-n)Fx\,.
\end{equation}
Provided $F>0$ and $x>0$ initially, the numerical simulations
show that $F$ continues to increase for $2 \le n <3$ 
without changing the sign of $x$. 
In the simulation of Fig.~\ref{backevo}, for example, 
the present value of $F$ is about 1.35 times as large as its initial value at the radiation 
era. This growth of $F$ is different from
$f(R)$ dark energy models in which the quantity 
$F=\partial f/\partial R$ decreases with time \cite{fRviable,fRviable2}.
Numerically we have also checked that the no-ghost condition 
$Q_s>0$ is satisfied from the radiation era to the dS epoch.
When $n=2$, Eq.~(\ref{eq:QdS}) shows that $Q_s \to \infty$
at the dS point. Hence the model proposed in Ref.~\cite{Silva}
may suffer from the lack of proper quantizations at the dS point.

In Fig.~\ref{csevofig} we plot the evolution of $c_s^2$ for
several different cases. When $n \neq 3$ the numerical 
simulations confirm the values $c_s^2 \simeq 5/6$ and
$c_s^2 \simeq 2/3$ during radiation and matter eras, 
respectively. Depending on the values of $n$ and $\omega$, 
the evolution of $c_s^2$ from the matter era to the dS epoch
is different. When $n=2.5$, for example, the propagation speed
remains subluminal ($c_s^2<1$) for $\omega>-85$ (as in the case (b) 
in Fig.~\ref{csevofig}), whereas it temporally reaches 
the superluminal regime ($c_s^2>1$) for $\omega<-85$.
The critical values of $\omega$ characterizing the border of 
subluminal and superluminal regimes decrease for smaller $n$.
In Fig.~\ref{omegafig}  the border line that marginally 
reaches the value $c_s^2=1$ is plotted as a dotted curve.
When $n=2$ the avoidance of the superluminal propagation 
corresponds to $\omega>-188$, see the case (a) in Fig.~\ref{csevofig}.
Note that this condition is consistent with the result 
found in Ref.~\cite{Silva}.

If $n$ is close to 3, the temporal superluminal propagation can 
be avoided only for the values of $\omega$ close to 0.
In particular, when $n=3$, we find that 
the superluminal propagation is inevitable for $\omega<0$.
When $n=3$, Eq.~(\ref{yestima}) shows that the variable $y$ 
{\it decreases} as $y \propto e^{-N}=a^{-1}$ unlike 
the cases with $n \neq 3$.
In the regime $x \ll y$, the variable $x$ increases as 
$x \propto e^{3N/2}=a^{3/2}$ during the radiation era. 
Hence, even if $x \ll y$ initially, there is an epoch at which 
$x$ and $y$ become the same order.
After the system reaches this epoch, the propagation speed 
(\ref{csra2}) is no longer valid. 
In the case (c) of Fig.~\ref{csevofig} we find that $c_s^2$
starts to evolve from the value $1/2$ as expected, 
but it enters the superluminal regime around the end 
of the radiation domination.
This peculiar evolution of $c_s^2$ is associated with 
the specific evolution of the variables $y$ and $x$ for $n=3$.

\section{Conclusions}
\label{conclude}

We have constructed modified gravitational models of dark energy 
starting from the general action (\ref{action}) without a field potential.
The presence of a non-linear self interaction 
$\xi(\phi) \square \phi (\partial^{\mu} \phi 
\partial_{\mu} \phi)$ allows us to recover the General Relativistic 
behavior in the regions of high density.
The functions $F(\phi)$, $B(\phi)$, and 
$\xi(\phi)$ are restricted in the power-law forms given 
in Eq.~(\ref{funchoice}) from the demand of 
obtaining dS solutions at late times.
The theory with $n=2$ corresponds to Brans-Dicke theory with 
the self-interaction $\xi(\phi) \propto \phi^{-2}$, which was 
recently studied in literature \cite{Silva}.
If $\xi(\phi)=0$, the theories with the functions 
$F(\phi)=M_{\rm pl}^2 (\phi/M_{\rm pl})^{3-n}$ and 
$B(\phi)=\omega ( \phi/M_{\rm pl})^{1-n}$
are equivalent to Brans-Dicke theory by introducing a new field 
$\chi \equiv F(\phi)$.
However, except for $n=2$, the presence of the field self interaction term
does not allow us to express theories with the functions
(\ref{funchoice}) as Brans Dicke theory plus
the term $\xi(\chi) \square \chi (\partial^{\mu} \chi 
\partial_{\mu} \chi)$ by such a field redefinition.

We have derived the dynamical equations (\ref{auto1})-(\ref{auto3})
to discuss the background cosmological evolution for the theories
with the functions (\ref{funchoice}). The evolution of the dimensionless
variables $x$ and $y$ during radiation and matter eras can be analytically
estimated, which is consistent with the results obtained by numerical 
integrations. The presence of these epochs demands the condition 
$n \le 3$. The variables $x$ and $y$ at the dS point, denoted as 
$x_{\rm dS}$ and $y_{\rm dS}$ respectively, are 
determined by solving Eqs.~(\ref{xdS}) and (\ref{ysolu}) 
for given $n$ and $\omega$. We also studied the stability of 
the dS point by considering the evolution of homogenous curvature 
perturbations and found that the stability condition 
is given by $(n-3)x_{\rm dS}<3$.

The viable model parameter space can be restricted further by 
studying conditions for the avoidance of ghosts and instabilities.
The no-ghost condition corresponds to $Q_s>0$, where $Q_s$
is defined in Eq.~(\ref{Qdef}).
Provided $F(\phi)>0$, we require that $x>0$ to avoid
ghosts during the cosmological evolution from the radiation 
era to the dS epoch. We note that the case $n=2$ is special 
because of the divergent behavior of the quantity $Q_s$ at the 
dS point. Interestingly the stability of the dS point is 
automatically satisfied for $n \le 3$ and $x_{\rm dS}>0$.

The instability of perturbations can be avoided for $c_s^2>0$, 
where the propagation speed squared $c_s$
is defined in Eq.~(\ref{cses}).
At the dS point $c_s^2$ reduces to Eq.~(\ref{eq:csdS}), 
which is positive for $n \ge 2$ under the conditions $n \le 3$
and $x_{\rm dS}>0$. Hence the viable parameter region of $n$
is constrained to be $2 \le n \le 3$. For the existence of the dS point 
the parameter $\omega$ is restricted in the range 
$\omega<-n(n-3)^2$ from Eq.~(\ref{xdS}).
When $n \neq 3$ the evolution of $c_s^2$ during radiation and matter 
dominated epochs can estimated as Eqs.~(\ref{csra}) and (\ref{csma}) 
respectively, which remain subluminal.
During the transition from the matter era to the dS epoch the propagation 
speed can be superluminal, depending on the values of $\omega$ and 
$n$. The avoidance of the temporal superluminal propagation gives 
the lower bound on $\omega$ for each value of $n$.
In Fig.~\ref{omegafig} we have plotted the viable model parameter space
and the region of the subluminal propagation in the $(n, \omega)$ plane.
When $n=3$ it is difficult to avoid the appearance of the superluminal mode
during the matter era because of the specific evolution of the variables $x$ and $y$.

It will be of interest to study the evolution of matter density perturbations and 
gravitational potentials to confront our theories with the observations of 
large scale structure, cosmic microwave background, and weak lensing.
We leave the detailed analysis of cosmological perturbations for future work.

\section*{ACKNOWLEDGEMENTS}
The work of A.\,D and S.\,T. was supported by the 
Grant-in-Aid for Scientific 
Research Fund of the JSPS Nos.~09314 and 30318802. 
S.\,T. also thanks financial support for the Grant-in-Aid for 
Scientific Research on Innovative Areas (No.~21111006).
We thank Reza Tavakol for his collaboration in the early stage 
of this work. We are grateful to M.~Sami for useful comments.

\appendix
\section{General de Sitter solutions}
\label{genede}

We will look for the existence of general dS solutions. 
Let us assume the following form for $\dot\phi$ on the dS
background ($H=H_{\rm dS}$):
\begin{equation}
\label{eq:xphi}
\dot\phi=H_{\rm dS}\,\phi\,x(\phi)\,.
\end{equation}
Once $x$ is a given function of $\phi$, Eq.~(\ref{eq:xphi}) 
can be easily solved as
\begin{equation}
\label{eq:perp}
\int_{\phi_0}^\phi\frac{\rd\tilde\phi}{\tilde\phi\,x(\tilde\phi)}
=H_{\rm dS}(t-t_0)\,,
\end{equation}
where $\phi_0$ and $H_{\rm dS}$ are constants.
Taking the time derivative of Eq.~(\ref{eq:xphi}), we find
\begin{equation}
\label{eq:x2phi}
\ddot\phi=H_{\rm dS}^2\phi\,x\,(x+\phi\,x_{,\phi})\, .
\end{equation}
We use the Friedmann equation (\ref{alg1}) 
to find $B$, that is
\begin{equation}
\label{eq:Bphi}
B=\frac{2 (3 F-6 \bar\xi\, \phi^3\, x^3+3 F_{,\phi}\, \phi\, x
+\bar\xi_{,\phi}\, \phi^4\, x^4)}{\phi^2\, x^2}\,,
\end{equation}
where $\bar\xi\equiv H_{\rm dS}^2\,\xi$. 
Combing this with Eq.~(\ref{alg2}), it follows that
\begin{equation}
\label{eq:elk2}
\bar\xi=\frac{6 F+5 F_{,\phi}\, \phi\, x+F_{,\phi\phi}\, 
\phi^2\, x^2+F_{,\phi}\, \phi\, x^2+F_{,\phi}\, \phi^2\, x\, x_{,\phi}}
{2\phi^3\, x^3\, (x+\phi\, x_{,\phi}+3)}\,.
\end{equation}
After giving $x$ and $F$ as functions of $\phi$, 
one finds $\xi$ and $B$ from Eqs.~(\ref{eq:Bphi}) and (\ref{eq:elk2}).
Thus we have proved the existence of dS solutions in terms of two free functions, 
$x(\phi)$ and $F(\phi)$. Since the field $\phi$ is known as a function of $t$ from 
Eq.~(\ref{eq:perp}), the evolution of $F$, $B$, $\xi$, $Q_s$, and $c_s^2$ 
at the dS point is also determined.
In particular, the evolution of the time-dependent homogeneous curvature perturbations 
about the dS background is given by 
\begin{equation}
\label{eq:solDES}
{\cal R}=c_1+c_2\int^\phi \rd\tilde\phi\left[\frac{\exp\{-3\int^{\tilde\phi }\rd\bar\phi\, [\bar\phi\, x(\bar\phi)]^{-1}\}}{Q_s(\tilde\phi)\, \tilde\phi\, x(\tilde\phi)}  \right]\,.
\end{equation}
If the second term on the r.h.s.\ of Eq.~(\ref{eq:solDES}) decays and/or oscillates with time, then the dS solution can be regarded as stable.

\section{Theories with $\lambda<0$}
\label{lamnega}

If $\lambda<0$, then the variable $y$ is negative from the 
definition (\ref{ydef}). For the existence of the matter fixed 
point (\ref{matter}) we require that $n>3$.
The no-ghost condition (\ref{xcon}) can be satisfied for
$x<0$. {}From Eq.~(\ref{lamre}) we have $\lambda<0$
for $-3<x_{\rm dS}<0$ and $n>3$. 
Meanwhile, when $n>3$, the propagation speed 
squared (\ref{eq:csdS}) is positive only in the region $x_{\rm dS}<-2(4+\sqrt{6})$, 
which is not compatible with the condition $-3<x_{\rm dS}<0$.

Hence the theories with $\lambda<0$ are not viable.


\end{document}